\documentclass[showpacs, showkeys,twocolumn,preprintnumbers,floatfix,amsmath,amssymb]{revtex4}
\usepackage{booktabs}
\usepackage{natbib}
\usepackage{amsmath,amsthm}
\usepackage{float}
\usepackage{graphicx}
\usepackage{epstopdf}
\usepackage{caption}
\usepackage{subfigure}
\usepackage{tabularx}
\usepackage{makecell}
\usepackage{threeparttable}
\usepackage{dcolumn}
\usepackage{bm}
\usepackage{color,epsfig,multirow}
\usepackage{array}
\usepackage{hyperref}
\usepackage{algorithm}
\usepackage{algorithmic}
\usepackage{threeparttable}
\usepackage{mathrsfs}

\DeclareMathOperator{\erf}{erf}
\DeclareMathOperator{\erfc}{erfc}

\newcommand{\cO}{\mathcal{O}}

\theoremstyle{definition}

\theoremstyle{remark}

\allowdisplaybreaks[4]

\begin{document}
\preprint{Preprint}

\title{Energy Stable Scheme for Random Batch Molecular Dynamics}
\author{Jiuyang Liang}\thanks{liangjiuyang@sjtu.edu.cn}
\author{Zhenli Xu}\thanks{xuzl@sjtu.edu.cn}
\author{Yue Zhao}\thanks{Current address: Department of Computational Mathematics, Science \& Engineering, Michigan State University, East Lansing, MI 48824, USA. Email: zhaoyu14@msu.edu}
\affiliation{School of Mathematical Sciences, CMA-Shanghai and MOE-LSC, Shanghai Jiao Tong University, Shanghai 200240, China}

\date{\today}
\begin{abstract}
The computational bottleneck of molecular dynamics is the pairwise additive long-range interactions between particles. The random batch Ewald (RBE) method provides a highly efficient and superscalable solver for long-range interactions, but the stochastic nature of this algorithm leads to unphysical self-heating effect during the simulation. We propose an energy stable scheme (ESS) for particle systems by employing a Berendsen-type energy bath. The scheme removes the notorious energy drift which exists due to the force error even when a symplectic integrator is employed. Combining the RBE with the ESS, the new method provides a perfect solution of the computational bottleneck of molecular dynamics at the microcanonical ensemble. Numerical results for primitive electrolyte and all-atom pure water systems demonstrate the attractive performance of the algorithm including its dramatically high accuracy, linear complexity and overcoming the energy drift for long-time simulations. 
\end{abstract}


\pacs{02.70.Ns, 87.15.Aa}
\keywords{Molecular dynamics simulations, long-range interactions, random batch Ewald method, energy conservation}

\maketitle

Molecular dynamics (MD) simulation is one of the most useful tools for studying structural and dynamical properties
of many-body systems in various areas  ~\cite{Axel1987Science,karplus1990molecular,Scott2018Neuron,Vesselin2002NatMater}.
The calculation of the force composed of pairwise additive interactions between particles is the most time-consuming part of the MD.
Particularly, the treatment of electrostatic forces is notoriously difficult due to the nature of long-range interactions. The random batch Ewald (RBE) method~\cite{Jin2020SISC} has emerged as a promising technique, outperforming the mainstream algorithms such as the particle-particle particle-mesh (PPPM) method~\cite{Hockney1988Computer} and solving the computational bottleneck of the MD simulation. 
The RBE introduces a stochastic approximation to the discrete Fourier transform instead of the communication-intensive fast Fourier transform (FFT) summation for most existing algorithms.
This technique achieves both linear scaling and superscalability in parallel computing, producing $1$ order of magnitude faster than the PPPM for large-scale simulations~\cite{liang2022superscalability}.
Moreover, the random batch method is successful in saving memory in constructing neighbor lists and accelerates short-range interactions~\cite{liang2021random2}.  

The stochastic nature of approximating the interacting force poses a challenge in maintaining energy conservation which is an important criterion for numerical performance.
The RBE introduces noise to Newton's equations of motion, resulting in an energy drift during the simulations, and symplectic algorithms for time integration cannot solve the problem.
This effect can be removed by using the Nos\'e-Hoover thermostat~\cite{hoover1985canonical} for the canonical ensemble and the Langevin barostat~\cite{liang2021random} for the isothermal-isobaric ensemble.
The energy conservation for the microcanonical (NVE) ensemble remains an open problem since it is devoid of thermal bath interference.
The NVE ensemble presents notable advantages for exploring adiabatic processes, phase transitions, and non-equilibrium dynamics~\cite{Frenkel2001Understanding,leimkuhler2015molecular}.
Note that in spite that the symplectic scheme \cite{FengQin2010, 10.1063/1.3623585,levesque1993molecular,toxvaerd2012energy,martyna1995symplectic} can be employed, the 
unphysical energy drift during long-time simulations has not been fully resolved~\cite{levitt1995potential,GRAZIANI2012105,hammonds2021shadow}
due to the error introduced in the force calculation. 

In this paper, we propose a novel approach for energy stable scheme (ESS) of MD simulations of particle systems in the NVE ensemble. This approach is particularly useful when the stochastic approximation is employed for the force calculation. We consider Coulomb systems with the RBE approximating the interacting force. The local random noise leads to energy drift, and the traditional velocity Verlet cannot eliminate it. The basic idea of the ESS is the principle of least local perturbation which was originally proposed by Berendsen \emph{et al.}~\cite{berendsen1984molecular} to approximate the energy variation arising in an ideal physical nonequilibrium experiment. The scheme is demonstrated to accurately reproduce the NVE results of both asymmetric electrolytes and all-atom bulk water systems, achieving an $O(\Delta t^2/P)$ energy fluctuation, with $P$ being the batch size of the RBE. The ESS allows us to produce accurate results by the stochastic approximation with a significant improvement in simulation efficiency.

To describe the details of the ESS approach, consider a charged system of $N$ particles in a cubic box of side length $L$ and volume $V=L^3$, with the $i$th particle having strength $q_i$, mass $m_i$ and position $\bm{r}_i$. The Hamiltonian of this system comprises the kinetic and potential energies
\begin{equation}
H=K(t)+U(t)
\end{equation}  
with $K=\sum_i m_{i}\bm{v}_{i}^{2}/2$ and the potential energy $U$ is represented by a combination of Lennard-Jones and long-range Coulomb interactions between particles. Under the periodic boundary condition, the Coulomb interaction among particles is usually evaluated by the Ewald-type summation~\cite{Ewald1921AnnPhys,hu2014infinite}, where the Coulomb kernel is split as the short- and long-range parts 
	$1/r=\erfc(\alpha r)/r+\erf(\alpha r)/r$
with $\alpha>0$ being a specified positive constant. The long-range component is smooth and slowly varying. The corresponding energy can be written as a Fourier series
\begin{equation}
U_{\text{coul}}^{\mathcal{F}}(\{\bm{r}_i\})=\frac{2\pi}{V}\sum_{\bm{k}\neq \bm{0}}\frac{1}{|\bm{k}|^2}|\rho(\bm{k})|^2 e^{-|\bm{k}|^2/4\alpha^2},
\end{equation}
with $\bm{k}=2\pi\bm{m}/L$, $\bm{m}\in{\mathbb{Z}^3}$, and the structure factor
\begin{equation}
	\rho(\bm{k}):=\sum_{i=1}^{N} q_i e^{i\bm{k}\cdot\bm{r}_i}.
\end{equation}
The long-range force acting on the $i$th particle is then given by $\bm{F}_{\text{coul}}^{\mathcal{F},i}=-\nabla_{\bm{r}_i}U_{\text{coul}}^{\mathcal{F}}(\{\bm{r}_i\})$, which can be calculated due to the rapid decay of the Fourier series. Tuning parameter $\alpha$ for improved efficiency and accelerating the discrete Fourier transform by the FFT, an $O(N\log N)$ complexity for the electrostatic interaction can be achieved~\cite{Hockney1988Computer,Darden1993JCP,essmann1995smooth}. Despite their efficiency, considerable energy drift during NVE ensemble simulations exists due to the nonsmoothness of the Ewald splitting  \cite{toxvaerd2012energy,hammonds2022optimization}. Other popular Coulomb solvers ~\cite{greengard1987fast,Barnes1986Nature,trottenberg2000multigrid,maggs2002local}  such as fast multipole methods also suffer from this issue~\cite{Arnold2013PRE}. It is remarked that various methods including shifted forces~\cite{toxvaerd2011communication} and kernel smoothing~\cite{hammonds2020shadow,hammonds2021shadow,Shamshirgar2019JCP} can mitigate the energy drift but they may also introduce system perturbations. For the PPPM, the increase of parameter $\alpha$ in Ewald splitting improves the accuracy and typically helps reduce the drift, but it requires a balance with the added computational cost.
 
In the RBE method~\cite{Jin2020SISC}, the discrete Fourier transform is calculated by importance sampling. One randomly takes $P$ Fourier modes $\{\bm{k}_{\ell}\}_{\ell=1}^{P}$ at each time step from the discrete Gaussian distribution $\mathscr{P}(\bm{k})=S^{-1}e^{-|\bm{k}|^2/4\alpha^2}$ with $S$ being the normalization factor. The long-range force is then approximated by
\begin{equation} \label{force}
	\widetilde{\bm{F}}_{\text{coul}}^{\mathcal{F},i}=-\frac{S}{P}\sum_{\ell=1}^{P}\frac{4\pi q_i\bm{k}_{\ell}}{V|\bm{k}_{\ell}|^2}\text{Im}\left(e^{-i\bm{k}_{\ell}\cdot\bm{r}_i}\rho(\bm{k}_{\ell})\right).
\end{equation}
The random minibatch strategy has its origin in the stochastic gradient descent~\cite{robbins1951stochastic}, firstly proposed for interacting particle systems~\cite{jin2020random}, and has succeeded in many areas~\cite{li2020random,carrillo2021consensus,ko2021model,golse2020random}. The convergence of this type of method is ensured due to the law of large numbers over time~\cite{jin2021convergence}. If $\alpha$ in the Ewald splitting has the same setup as the PPPM and $P$ is independent of $N$, the complexity of the RBE method is $O(N)$ and $O(1)$ communication cost for parallel computing, significantly outperforming the FFT-based method.

However, the stochastic nature of the RBE poses a challenge to energy conservation. To handle this issue, consider the classical symplectic velocity-Verlet (VV) scheme~\cite{Frenkel2001Understanding,verlet1967computer} employed in an NVE-ensemble simulation  
\begin{equation}\label{eq::VV}
	\begin{aligned}
		&\bm{v}_i(t+\Delta t/2)=\bm{v}_i(t)+\frac{\Delta t}{2m_i}\bm{F}_i(t),\\[0.5em]
		&\bm{r}_i(t+\Delta t)=\bm{r}_i(t)+\bm{v}_i(t+\Delta t/2) \Delta t,\\[0.5em]
		&\bm{v}_i(t+\Delta t)=\bm{v}_i(t+\Delta t/2)+\frac{\Delta t}{2m_i}\bm{F}_i(t+\Delta t).
	\end{aligned}
\end{equation}
It was shown that the VV scheme conserves the  ``Shadow Hamiltonian''~\cite{toxvaerd1994hamiltonians} by the backward analysis~\cite{griffiths1986scope,hairer1993backward}. If $\bm{F}_{i}$ is perturbated by $\bm{\delta}_i$, for the RBE, which is $\bm{\delta}_{i}=\widetilde{\bm{F}}_{\text{coul}}^{\mathcal{F},i}-\bm{F}_{\text{coul}}^{\mathcal{F},i}$, the Hamitonian can be expanded in the following form~\cite{toxvaerd2012energy} 
\begin{equation}\label{eq::4212}
	\begin{split}
		H(t_{n+1})&=H_{0}+\frac{\Delta t}{2}\sum_{j=0}^{n}\sum_{i=1}^{N}\bm{v}_i^T(t_{j+\frac{1}{2}})\left[\bm{\delta}_i(t_{j+1})+\bm{\delta}_i(t_j)\right]\\
		&+O(\Delta t^2),
	\end{split}
\end{equation}
where $H_{0}=H(t_0)$ is the initial Hamiltonian.  Eq.~\eqref{eq::4212} reveals that the cumulative error for the Hamiltonian stems from the sum of uncorrelated random projections aligned with the corresponding velocity direction $\bm{e}_{\bm{v}_i}$.
The random noise for the RBE has a diffusive-like effect, leading to energy instability which cannot be fixed by reducing the time step.

We introduce a weakly-coupled external energy bath on the VV scheme to describe the evolution of the kinetic energy,
\begin{equation}\label{eq::Breendsen2}
	\begin{split}
		&\frac{d}{dt}K(t)=\frac{1}{\gamma}\left(H_0-\widetilde{H}(t)\right),
	\end{split}
\end{equation}
where $\gamma$ is a coupling parameter, and $\widetilde{H}$ represents the approximate Hamiltonian due to the use of the RBE, 
\begin{equation}\label{eq::randomHamilton}
	\widetilde{H}=K+U+\widetilde{U}_{\text{coul}}^{\mathcal{F}}-U_{\text{coul}}^{\mathcal{F}}.
\end{equation}
Here the approximation for the long-range part of Coulomb energy reads
\begin{equation}
	\widetilde{U}_{\text{coul}}^{\mathcal{F}}=\frac{2\pi S}{PV}\sum_{\ell=1}^{P}\frac{1}{|\bm{k}_{\ell}|^2}|\rho(\bm{k}_{\ell})|^2
\end{equation}
where the frequencies in batch $\{\bm{k}_{\ell}\}_{\ell=1}^{P}$ are the same as those in Eq.~\eqref{force} and thus
\begin{equation}\label{eq::condition}
	-\nabla_{\bm{r}_i}\widetilde{U}_{\text{coul}}^{\mathcal{F}}(\{\bm{r}_i\})=\widetilde{\bm{F}}_{\text{coul}}^{\mathcal{F},i}
\end{equation}
in each time step. The evolution employing the RBE force is equivalent to the Hamiltonian dynamics based on $\widetilde{H}$ which is an unbiased estimator of $H$. One remarks that the calculation of $\widetilde{H}$ is not costly since $U-U_{\text{coul}}^{\mathcal{F}}$ in Eq.~\eqref{eq::randomHamilton} is just the short-range interactions including the Lennard-Jones contribution and those from the Coulomb potential. 

The parameter $\gamma$ in Eq.~\eqref{eq::Breendsen2} acts as a relaxation time dictating the interval between successive dissipation of artificial heat, driving the dynamics to the correct NVE ensemble.  
A large $\gamma$ implies a small dissipation of artificial heat in the system, leading to the potential persistence of substantial energy drifts. When $\gamma\rightarrow\infty$, the dynamics will be consistent with the classical NVE ensemble.
Conversely, if $\gamma$ is too small, the velocity fluctuation will be nonphysically large. Our systematic tests illustrate that $\gamma$ in the range of $10-100~\Delta t$ produces correct results.

With the energy bath, one introduces an additional term into the equations of motion, which is proportional to the velocity. 
One obtains
\begin{equation}
	d\bm{v}_i=\frac{\bm{F}_i}{m_i}dt+\lambda\bm{v}_idt,
\end{equation}
where $\lambda=(H_0-{\widetilde{H}})/(2\gamma K)$ is the solution to Eq.~\eqref{eq::Breendsen2}. This equation can be solved by first employing the VV scheme \eqref{eq::VV} for the positions and velocities, followed by a proportional scaling of the velocities from $\bm{v}_i$ to $\xi \bm{v}_i$ with
\begin{equation}\label{eq::factor}
	\xi=\left[1+\dfrac{\Delta t }{\gamma K}(H_0-\widetilde{H})\right]^{1/2}.
\end{equation}
One sees that Eq.~\eqref{eq::Breendsen2} does not add any effect when $\widetilde{H}=H_0$, and will adaptively dissipate artificial heat while $\widetilde{H}\neq H_0$. The dissipation occurs at each time step along the direction of $\bm{v}_i$ of every particle that is consistent with Eq.~\eqref{eq::4212}. Note that the same damping factor $\xi$ is used for all particles to maintain the conservation of momentum.

The ESS algorithm is summarized by the following steps. For a given $\Delta t$, one first samples $P$ Fourier frequencies from discrete distribution $\mathscr{P}(\bm{k})$ and evolves the VV dynamics Eq.~\eqref{eq::VV} using the RBE force $\widetilde{\bm{F}}_{\text{coul}}^{\mathcal{F},i}$. 
Subsequently, one scales the velocity $\bm{v}_i$ to $\xi\bm{v}_i$ according to Eq.~\eqref{eq::factor}, in which the random Hamiltonian $\widetilde{H}$ is used. One proceeds with the loop of these two steps to conduct the simulation until specific termination conditions are fulfilled. 
In each step, identical samples are used approximate both Hamiltonian and forces. This significantly reduces communication rounds. The computation remains $O(N)$ with a slight increase in cost, making it almost as efficient as the original RBE \cite{liang2022superscalability}. The proposed ESS can be readily extended to other electrostatic algorithms~\cite{liang2021random2,liang2023SISC,gao2023screening}. For instance, in cases where electrostatic correlations are significant, the symmetry-preserving RBE~\cite{gao2023screening} can greatly enhance the accuracy of the RBE. By incorporating the ESS, this method can further improve accuracy due to the energy stability. Additionally, the scheme can be applied to other interacting kernels, including dipolar potentials~\cite{PhysRevLett.119.155501}, dispersion potentials~\cite{10.1063/1.2770730}, and screened Coulomb potentials~\cite{PhysRevLett.88.065002} commonly used in plasma simulations.

The performance of the ESS is validated through MD simulations of a primitive electrolyte and an all-atom water systems by measuring  radial distribution functions (RDF), mean-square displacements (MSD), and average energy fluctuations per atom over extended simulations ($\langle H(t)-H_0\rangle/N$). Simulations were done in LAMMPS~\cite{thompson2022lammps} on the ``Siyuan Mark-I'' cluster by comparing our method to the VV scheme with the shifted PPPM~\cite{LEVITT1995215} as the Coulomb solver. The parameter $\alpha$ in the Ewald splitting is tuned to ensure an error level of $10^{-4}$~\cite{deserno1998mesh} for the PPPM. The same $\alpha$ is also used for the ESS. The ``Reference'' solution is obtained by the results of a small step size ensuring energy conservation. The ESS results of different $\gamma$ are then compared with the PPPM and the ``Reference'' solutions to demonstrate the effectiveness of our scheme.

\begin{figure*}[t!]	
	\centering
	\includegraphics[width=0.95\textwidth]{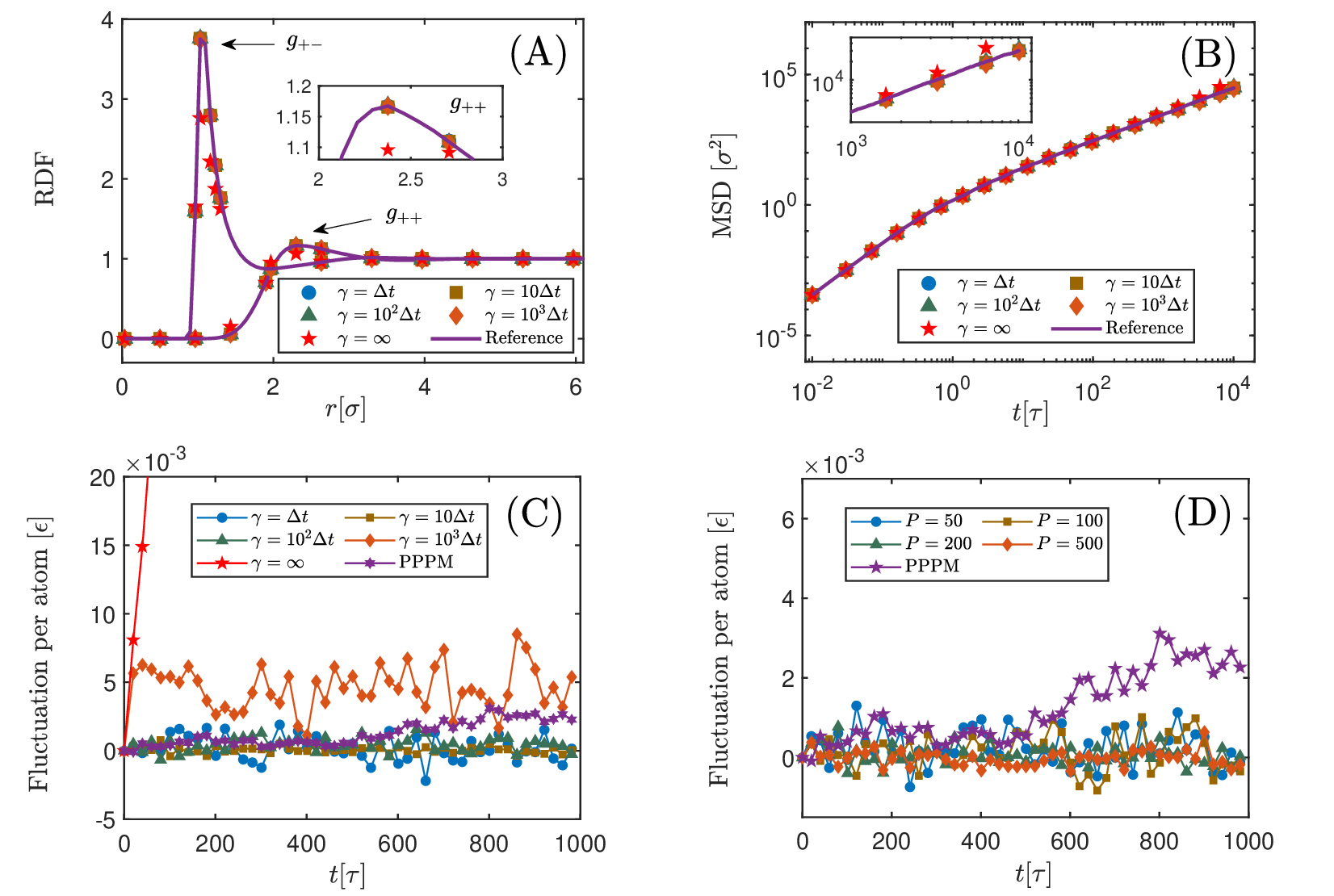}
	\caption{
	Results of the electrolyte systems. (A-C) The cation-anion ($g_{+-}$) and cation-cation ($g_{++}$) RDFs, the MSD of cations and the energy fluctuation for different values of $\gamma$ with a fixed batch size of $P=500$; (D) Energy fluctuation for different $P$ with fixed $\gamma=10\Delta t$. The results obtained using the PPPM method are also included in (C) and (D) for comparison.
	} 
	\label{fig:gamma}
\end{figure*}

The first benchmark is an asymmetric primitive electrolyte system with $3:1$ salt~\cite{liang2022hsma} in a $L=25\sigma$ box, with the ions being modeled as soft spheres of diameter $\sigma$. The system contains $750$ cations and $2250$ anions (volume fraction of $10.05\%$), interacting through a repulsive shifted-truncated Lennard-Jones potential with an energy coupling $\varepsilon_{\text{LJ}}=k_{\text{B}}T$, where $k_{\text{B}}$ is the Boltzmann constant. The continuum solvent is characterized by a Bjerrum length $\ell_{\text{B}}=3.5\sigma$. We use a time step of $\Delta t=0.01\tau$, with a real-space cutoff of $8\sigma$ for both the ESS and PPPM.
Figure~\ref{fig:gamma}(A-C)  displays the results with varying $\gamma$ for a fixed batch size $P=500$. When $\gamma=\infty$ (without the bath), a significant energy drift can be observed, leading to a simulation blowup after $6\times10^{5}$ steps. In this case,  $g_{+-}$, which is the RDF between trivalent cations and monovalent anions, deviates significantly from the reference values, revealing that the original RBE cannot accurately capture the strong Coulomb coupling effect without the bath. For finite $\gamma$, the RDF and MSD curves almost overlap. The heat dissipation increases as $\gamma$ decreases, and the energy oscillation is pretty stable with $\gamma\leq100\Delta t$. Notably, the PPPM-based simulations show significant energy drift due to the nonsmoothness of the Ewald decomposition splitting~\cite{DEShaw2020JCP,hammonds2022optimization}. Further calculation on varying $P$ with $\gamma=10\Delta t$ shows that the energy fluctuations are reduced with the increase of the batch size (Figure~\ref{fig:gamma}(D)). One can see that even at a small size $P=50$, the ESS outperforms the PPPM without the bath. One can expect that the advantages of the ESS will become evident for systems with stronger Coulomb coupling due to that the energy bath can adaptively dissipate the additional noise.

\begin{figure*}[!ht]	
	\centering
	\includegraphics[width=0.95\textwidth]{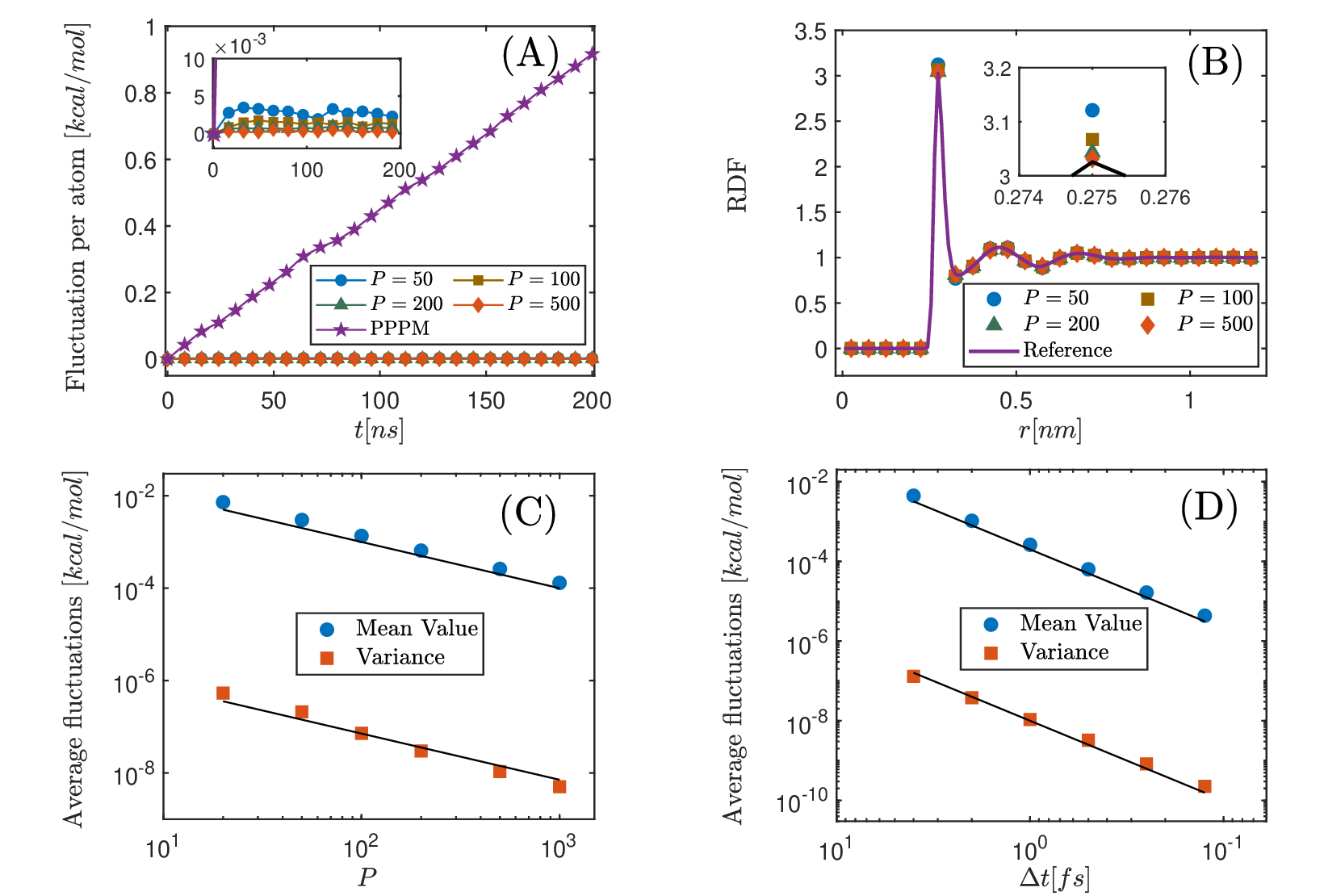}
	\caption{
		Results of the water system. (A-B) Energy fluctuation per atom and RDF of oxygen-oxygen pairs for different values of $P$ with a fixed damping factor of $\gamma=10\Delta t$; (C-D) Log-log plots of mean value and variance of energy fluctuation. The solid lines in (C) and (D) indicate rates of $\cO(1/P)$ and $\cO(\Delta t^2)$, respectively.
	}
	\label{fig:WaterSPCE}
\end{figure*}

The second example is an all-atom SPC/E bulk water system~\cite{berendsen1987missing} with $21624$ atoms in a cubic box of dimensions $6\,nm$. Energy fluctuations per atom during $1\,\mu s$ simulations, with an integration step of $\Delta t=1\,fs$ and $\gamma=10 \Delta t$, were analyzed for different batch sizes (Fig.~\ref{fig:WaterSPCE}(A)). The PPPM results exhibit a significant drift due to the lack of smoothness in Ewald splitting, which increases linearly over time. Oppositely, the ESS results perform well in energy stability for all $P$ from 50 to 500. The RDF results in Fig.~\ref{fig:WaterSPCE}(B) show the convergence of the results with the increase of $P$ (see the peaks in the enlarged subplot). Fig.~\ref{fig:WaterSPCE}(C-D) presents the convergence rate of the ESS for results of the ensemble average of energy fluctuation with varying  $P$ and  $\Delta t$, respectively. These results clearly display the $O(1/P)$ and $O(\Delta t^2)$ rates converging to $H_0$, namely, the Hamitonian satisfies $H(t_n)=H_0+O\left(\Delta t^2/P\right)$. Interestingly, the second order fluctuation is similar to the shadow Hamiltonian in symplectic VV dynamics~\cite{toxvaerd1994hamiltonians,leimkuhler2004simulating}.

\begin{figure}[ht]	
	\centering
	\includegraphics[width=0.48\textwidth]{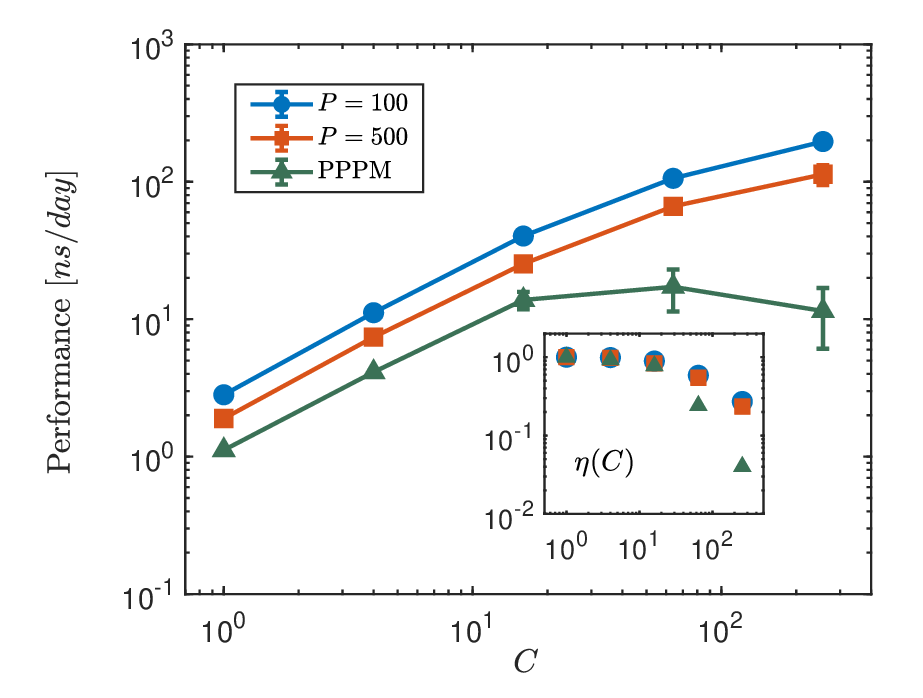}
	\caption{The comparison of CPU performance versus core number $C$ for ESS with batch sizes $P=100,~500$ and PPPM. The inset illustrates strong parallel scalability.}
	\label{fig:TimeScaling}
\end{figure}

The combination of the ESS and RBE offers high-accuracy results, whilt it also significantly reduces computational cost and enhances parallel scalability for the long-range electrostatic force. To assess the CPU performance, Fig.~\ref{fig:TimeScaling} presents these results using the same SPC/E bulk water system as mentioned earlier, with a time step of $1\,fs$. Clearly, the ESS notably outperforms the PPPM, and with an increse in CPU cores up to $C=256$, a tenfold improvement in performance ($ns$/day) can be achieved, owing to the superscalability of the RBE~\cite{liang2022superscalability}. Our additional tests on the mean and variance of average energy demonstrate that the ESS method with $P=500$ and a time step of $1\,fs$ achieves nearly the same accuracy as the PPPM method with a time step of $0.1\,fs$ for $2\,ns$ simulations and $0.01\,fs$ for $0.2\,\mu s$ simulations. This indicates that the ESS can further improve by $1-2$ orders of magnitude with a smaller time step. We remark that a fair comparison is difficult considering the difference in code optimization, but our results clearly demonstrate the attractive performance of the ESS scheme.

In summary, we have developed an efficient ESS for the random batch molecular dynamics, which shows promise in achieving both accuracy and efficiency in simulations. The ESS maintains the same $O(N)$ complexity and superscalability as the original RBE, but with an additional advantage of conserving long-term energy stability. The fluctuation rate is comparable to symplectic methods employing exact Hamiltonian with a small price of the computational cost. The ESS can be straightforwardly extended to many other interacting systems, particularly, the systems of short-range interactions where the random batch method is used to construct the neighbor list ~\cite{liang2021random2}. These will be the topic of our ongoing work.

\section*{Acknowledgement}
This work is supported by the National Natural Science Foundation of China (grant Nos. 12325113 and 12071288) and the Science and Technology Commission of Shanghai Municipality (grant No. 21JC1403700).
The authors also acknowledge the support from the HPC center of Shanghai Jiao Tong University. 

\section*{Conflict of interest}
The authors declare that they have no conflict of interest.

\section*{Data Availability Statement}
The data that support the findings of this study are available from the corresponding author upon reasonable request.


\begin{thebibliography}{10}
\expandafter\ifx\csname url\endcsname\relax
  \def\url#1{\texttt{#1}}\fi
\expandafter\ifx\csname urlprefix\endcsname\relax\def\urlprefix{URL }\fi

\bibitem{Axel1987Science}
A.~T. Br\"unger, J.~Kuriyan, M.~Karplus, {Crystallographic R factor refinement
  by molecular dynamics}, Science 235 (1987) 458--460.

\bibitem{karplus1990molecular}
M.~Karplus, G.~A. Petsko, Molecular dynamics simulations in biology, Nature 347
  (1990) 631--639.

\bibitem{Scott2018Neuron}
S.~A. Hollingsworth, R.~O. Dror, Molecular dynamics simulation for all, Neuron
  99~(6) (2018) 1129--1143.

\bibitem{Vesselin2002NatMater}
V.~Yamakov, D.~Wolf, S.~R. Phillpot, A.~K. Mukherjee, H.~Gleiter, Dislocation
  processes in the deformation of nanocrystalline aluminium by
  molecular-dynamics simulation, Nat. Mater. 1~(1) (2002) 45--49.

\bibitem{Jin2020SISC}
S.~Jin, L.~Li, Z.~Xu, Y.~Zhao, {A random batch Ewald method for particle
  systems with Coulomb interactions}, SIAM J. Sci. Comput. 43 (2021)
  B937--B960.

\bibitem{Hockney1988Computer}
R.~W. Hockney, J.~W. Eastwood, {Computer Simulation Using Particles}, CRC
  Press, 1988.

\bibitem{liang2022superscalability}
J.~Liang, P.~Tan, Y.~Zhao, L.~Li, S.~Jin, L.~Hong, Z.~Xu, {Superscalability of
  the random batch Ewald method}, J. Chem. Phys. 156~(1) (2022) 014114.

\bibitem{liang2021random2}
J.~Liang, Z.~Xu, Y.~Zhao, Random-batch list algorithm for short-range molecular
  dynamics simulations, J. Chem. Phys. 155~(4) (2021) 044108.

\bibitem{hoover1985canonical}
W.~G. Hoover, {Canonical dynamics: Equilibrium phase-space distributions},
  Phys. Rev. A 31~(3) (1985) 1695.

\bibitem{liang2021random}
J.~Liang, P.~Tan, L.~Hong, S.~Jin, Z.~Xu, L.~Li, {A random batch Ewald method
  for charged particles in the isothermal--isobaric ensemble}, J. Chem. Phys.
  157~(14) (2022) 144102.

\bibitem{Frenkel2001Understanding}
D.~Frenkel, B.~Smit, {Understanding Molecular Simulation: From Algorithms to
  Applications}, Vol.~1, Elsevier, 2001.

\bibitem{leimkuhler2015molecular}
B.~Leimkuhler, C.~Matthews, {Molecular Dynamics}, Interdisciplinary Applied
  Mathematics, Springer, 2015.

\bibitem{FengQin2010}
K.~Feng, M.~Qin, Symplectic Geometric Algorithms for Hamiltonian Systems,
  Zhejiang Publishing United Group and Springer, 2010.

\bibitem{10.1063/1.3623585}
T.~S. Ingebrigtsen, S.~Toxvaerd, O.~J. Heilmann, T.~B. Schrøder, J.~C. Dyre,
  {{NVU dynamics. I. Geodesic motion on the constant-potential-energy
  hypersurface}}, J. Chem. Phys. 135~(10) (2011) 104101.

\bibitem{levesque1993molecular}
D.~Levesque, L.~Verlet, Molecular dynamics and time reversibility, J. Stat.
  Phys. 72 (1993) 519--537.

\bibitem{toxvaerd2012energy}
S.~Toxvaerd, O.~J. Heilmann, J.~C. Dyre, {Energy conservation in molecular
  dynamics simulations of classical systems}, J. Chem. Phys. 136~(22) (2012)
  224106.

\bibitem{martyna1995symplectic}
G.~J. Martyna, M.~E. Tuckerman, {Symplectic reversible integrators:
  Predictor--corrector methods}, J. Chem. Phys. 102~(20) (1995) 8071--8077.

\bibitem{levitt1995potential}
M.~Levitt, M.~Hirshberg, R.~Sharon, V.~Daggett, Potential energy function and
  parameters for simulations of the molecular dynamics of proteins and nucleic
  acids in solution, Comput. Phys. Commun. 91~(1-3) (1995) 215--231.

\bibitem{GRAZIANI2012105}
F.~R. Graziani, V.~S. Batista, L.~X. Benedict, J.~I. Castor, H.~Chen, S.~N.
  Chen, C.~A. Fichtl, J.~N. Glosli, P.~E. Grabowski, A.~T. Graf, S.~P.
  Hau-Riege, A.~U. Hazi, S.~A. Khairallah, L.~Krauss, A.~B. Langdon, R.~A.
  London, A.~Markmann, M.~S. Murillo, D.~F. Richards, H.~A. Scott, R.~Shepherd,
  L.~G. Stanton, F.~H. Streitz, M.~P. Surh, J.~C. Weisheit, H.~D. Whitley,
  {Large-scale molecular dynamics simulations of dense plasmas: The Cimarron
  Project}, High Energy Density Phys. 8~(1) (2012) 105--131.

\bibitem{hammonds2021shadow}
K.~D. Hammonds, D.~M. Heyes, {Shadow Hamiltonian in classical NVE molecular
  dynamics simulations involving Coulomb interactions}, J. Chem. Phys. 154~(17)
  (2021) 174102.

\bibitem{berendsen1984molecular}
H.~J.~C. Berendsen, J.~P.~M. Postma, W.~F. van Gunsteren, A.~DiNola, J.~R.
  Haak, {Molecular dynamics with coupling to an external bath}, J. Chem. Phys.
  81~(8) (1984) 3684--3690.

\bibitem{Ewald1921AnnPhys}
P.~P. Ewald, {Die Berechnung optischer und elektrostatischer Gitterpotentiale},
  Ann. Phys. 369 (1921) 253--287.

\bibitem{hu2014infinite}
Z.~Hu, {Infinite boundary terms of Ewald sums and pairwise interactions for
  electrostatics in bulk and at interfaces}, J. Chem. Theory Comput. 10 (2014)
  5254--5264.

\bibitem{Darden1993JCP}
T.~Darden, D.~York, L.~Pedersen, {Particle mesh Ewald: An $N\log(N)$ method for
  Ewald sums in large systems}, J. Chem. Phys. 98 (1993) 10089--10092.

\bibitem{essmann1995smooth}
U.~Essmann, L.~Perera, M.~L. Berkowitz, T.~Darden, H.~Lee, L.~G. Pedersen, {A
  smooth particle mesh Ewald method}, J. Chem. Phys. 103 (1995) 8577--8593.

\bibitem{hammonds2022optimization}
K.~D. Hammonds, D.~M. Heyes, {Optimization of the Ewald method for calculating
  Coulomb interactions in molecular simulations}, J. Chem. Phys. 157~(7) (2022)
  074108.

\bibitem{greengard1987fast}
L.~Greengard, V.~Rokhlin, A fast algorithm for particle simulations, {J.
  Comput. Phys.} 73 (1987) 325--348.

\bibitem{Barnes1986Nature}
J.~Barnes, P.~Hut, {A hierarchical $O(N\log N)$ force-calculation algorithm},
  Nature 324 (1986) 446--449.

\bibitem{trottenberg2000multigrid}
U.~Trottenberg, C.~W. Oosterlee, A.~Schuller, Multigrid, Elsevier, 2000.

\bibitem{maggs2002local}
A.~Maggs, V.~Rossetto, {Local simulation algorithms for Coulomb interactions},
  Phys. Rev. Lett. 88~(19) (2002) 196402.

\bibitem{Arnold2013PRE}
A.~Arnold, F.~Fahrenberger, C.~Holm, O.~Lenz, M.~Bolten, H.~Dachsel, R.~Halver,
  I.~Kabadshow, F.~G\"ahler, F.~Heber, Comparison of scalable fast methods for
  long-range interactions, Phys. Rev. E 88~(6) (2013) 063308.

\bibitem{toxvaerd2011communication}
S.~Toxvaerd, J.~C. Dyre, {Communication: Shifted forces in molecular dynamics},
  J. Chem. Phys. 134~(8) (2011) 081102.

\bibitem{hammonds2020shadow}
K.~D. Hammonds, D.~M. Heyes, {Shadow Hamiltonian in classical NVE molecular
  dynamics simulations: A path to long time stability}, J. Chem. Phys. 152~(2)
  (2020) 024114.

\bibitem{Shamshirgar2019JCP}
D.~S. Shamshirgar, R.~Yokota, A.-K. Tornberg, B.~Hess, {Regularizing the fast
  multipole method for use in molecular simulation}, J. Chem. Phys. 151 (2019)
  234113.

\bibitem{robbins1951stochastic}
H.~Robbins, S.~Monro, A stochastic approximation method, Ann. Math. Stat.
  (1951) 400--407.

\bibitem{jin2020random}
S.~Jin, L.~Li, J.-G. Liu, {Random batch methods (RBM) for interacting particle
  systems}, J. Comput. Phys. 400 (2020) 108877.

\bibitem{li2020random}
L.~Li, Z.~Xu, Y.~Zhao, {A random-batch Monte Carlo method for many-body systems
  with singular kernels}, SIAM J. Sci. Comput. 42~(3) (2020) A1486--A1509.

\bibitem{carrillo2021consensus}
J.~A. Carrillo, S.~Jin, L.~Li, Y.~Zhu, A consensus-based global optimization
  method for high dimensional machine learning problems, ESAIM Contr. Optim.
  Ca. 27 (2021) S5.

\bibitem{ko2021model}
D.~Ko, E.~Zuazua, Model predictive control with random batch methods for a
  guiding problem, Math. Models Methods Appl. Sci. 31~(08) (2021) 1569--1592.

\bibitem{golse2020random}
F.~Golse, S.~Jin, T.~Paul, {The random batch method for $N$-body quantum
  dynamics}, J. Comput. Math. 39~(6) (2021) 897--922.

\bibitem{jin2021convergence}
S.~Jin, L.~Li, J.-G. Liu, Convergence of the random batch method for
  interacting particles with disparate species and weights, SIAM J. Numer.
  Anal. 59~(2) (2021) 746--768.

\bibitem{verlet1967computer}
L.~Verlet, {Computer ``experiments'' on classical fluids. I. Thermodynamical
  properties of Lennard-Jones molecules}, Phys. Rev. 159~(1) (1967) 98.

\bibitem{toxvaerd1994hamiltonians}
S.~Toxvaerd, Hamiltonians for discrete dynamics, Phys. Rev. E 50~(3) (1994)
  2271.

\bibitem{griffiths1986scope}
D.~F. Griffiths, J.~M. Sanz-Serna, On the scope of the method of modified
  equations, SIAM J. Sci. Stat. Comput. 7~(3) (1986) 994--1008.

\bibitem{hairer1993backward}
E.~Hairer, Backward analysis of numerical integrators and symplectic methods,
  Annals Num. Math. 1 (1994) 107--132.

\bibitem{liang2023SISC}
J.~Liang, Z.~Xu, Q.~Zhou, {Random batch sum-of-Gaussians method for molecular
  dynamics simulations of particle systems}, SIAM J. Sci. Comput. 45~(5) (2023)
  B591--B617.

\bibitem{gao2023screening}
W.~Gao, Z.~Hu, Z.~Xu, A screening condition imposed stochastic approximation
  for long-range electrostatic correlations, J. Chem. Theory Comput. 19~(15)
  (2023) 4822--4827.

\bibitem{PhysRevLett.119.155501}
L.~Spiteri, R.~Messina, Dipolar crystals: The crucial role of the
  clinohexagonal prism phase, Phys. Rev. Lett. 119 (2017) 155501.

\bibitem{10.1063/1.2770730}
P.~J. in~'t Veld, A.~E. Ismail, G.~S. Grest, {{Application of Ewald summations
  to long-range dispersion forces}}, J. Chem. Phys. 127 (2007) 144711.

\bibitem{PhysRevLett.88.065002}
G.~Salin, J.-M. Caillol, Transport coefficients of the {Y}ukawa one-component
  plasma, Phys. Rev. Lett. 88 (2002) 065002.

\bibitem{thompson2022lammps}
A.~P. Thompson, H.~M. Aktulga, R.~Berger, D.~S. Bolintineanu, W.~M. Brown,
  P.~S. Crozier, P.~J. in't Veld, A.~Kohlmeyer, S.~G. Moore, T.~D. Nguyen,
  et~al., {LAMMPS-a flexible simulation tool for particle-based materials
  modeling at the atomic, meso, and continuum scales}, Comput. Phys. Commun.
  271 (2022) 108171.

\bibitem{LEVITT1995215}
M.~Levitt, M.~Hirshberg, R.~Sharon, V.~Daggett, Potential energy function and
  parameters for simulations of the molecular dynamics of proteins and nucleic
  acids in solution, Comput. Phys. Commun. 91~(1) (1995) 215--231.

\bibitem{deserno1998mesh}
M.~Deserno, C.~Holm, {How to mesh up Ewald sums. II. An accurate error estimate
  for the particle--particle--particle-mesh algorithm}, J. Chem. Phys. 109~(18)
  (1998) 7694--7701.

\bibitem{liang2022hsma}
J.~Liang, J.~Yuan, Z.~Xu, {HSMA: An $O(N)$ electrostatics package implemented
  in LAMMPS}, Comput. Phys. Commun. 276 (2022) 108332.

\bibitem{DEShaw2020JCP}
C.~Predescu, A.~K. Lerer, R.~A. Lippert, B.~Towles, J.~P. Grossman, R.~M.
  Dirks, D.~E. Shaw, {The u-series: A separable decomposition for
  electrostatics computation with improved accuracy}, J. Chem. Phys. 152 (2020)
  084113.

\bibitem{berendsen1987missing}
H.~J.~C. Berendsen, J.~R. Grigera, T.~P. Straatsma, The missing term in
  effective pair potentials, J. Phys. Chem. 91~(24) (1987) 6269--6271.

\bibitem{leimkuhler2004simulating}
B.~Leimkuhler, S.~Reich, Simulating Hamiltonian Dynamics, {Cambridge University
  Press}, 2005.

\end{thebibliography}

\end{document}